\documentclass[twoside,11pt]{article}

\usepackage[cp1251]{inputenc}

\usepackage{graphicx}

\usepackage{amssymb}
\usepackage{amsmath}
\usepackage{amsfonts}
\usepackage{wasysym}
\usepackage{revsymb}
\usepackage{amsbsy}
\usepackage{bm}

\usepackage{titlesec}

\usepackage{cite}

\titleformat*{\section}{\bfseries}
\titleformat*{\subsection}{\bfseries}

\begin{document}           % End of preamble and beginning of text.

\title{\Large 
On the conditions for the classicality of a quantum particle}
\author{V.E. Kuzmichev, V.V. Kuzmichev\\[0.5cm]
\itshape Bogolyubov Institute for Theoretical Physics,\\
\itshape National Academy of Sciences of Ukraine, Kiev, 03143 Ukraine}

\date{}

\maketitle

\begin{abstract}
Conditions under which a quantum particle is described using classical quantities are studied. The one-dimensional 
(1D) and three-dimensional (3D) problems are considered. It is shown that the sum of the contributions from all 
quantum corrections (in the WKB sense) strictly vanishes, when a quantum particle interacts with some specific medium. 
The indices of refraction of such media are found. In this case, the smallness of the Planck constant is not assumed.
The momenta of quantum particles in these media and the wave functions of stationary states are determined. It is found that, for the 1D case,
the wave function is similar to that of the test particle with zero binding energy in a singular attractive potential, which admits ``fall'' to the center.
For the 3D case with central symmetry, a stationary state, describing a quantum particle with a classical momentum, is defined by the wave function,
which has the resonance of width about two de Broglie wavelengths.
\end{abstract}

PACS numbers: 03.65.-w, 02.30.Mv, 03.65.Sq, 03.65.Ta 

\section{Introduction}\label{sec:1}
Despite the fact that, by now, the mathematical apparatuses of nonrelativistic and relativistic (field) quantum theories have been sufficiently developed
with regard to their practical application to explain different physical phenomena, some fundamental questions about the relationship between quantum 
theory and classical physics still attract intense attention of physicists and mathematicians \cite{Ros86,tHo}.

In the present paper, the conditions under which a quantum particle has a classical momentum, while retaining its wave properties, are studied. 
By quantum particle, we mean a particle in a given force (potential) field, which is described by the Schr\"{o}dinger equation. This equation can be 
reduced to the generalized (quantum) Hamilton-Jacobi equation for the phase of the wave function \cite{Mes,Mad,Bohm}. The latter equation differs from the 
Hamilton-Jacobi equation of classical mechanics in that it contains a term nonlinear with respect to the derivatives of the phase and proportional 
to $\hbar^{2}$, which takes into account quantum effects. This term is often referred to as the ``quantum potential'' \cite{Bohm,Bohm2,Tak}.
Its influence on the motion of a quantum particle has been studied in various model systems, bringing the insights into the relations between classical and 
quantum physics (see, e.g., Refs. \cite{Ros64a,Ros64b,Ros65,Ros74,Phi,Bow,Den,Hei15,Hei16,TseHei,Per,Hoj20a,Hoj20b,Hoj20c} and references therein).

The WKB method, as is known, consists in expanding a nonlinear part of the generalized Hamilton-Jacobi 
equation in a series in powers of the Planck constant $\hbar$, followed by equating the terms of the series with the same powers of  $\hbar$.
The practical application of this method of solving the Schr\"{o}dinger equation is limited by the conditions imposed on the behavior of the potential
(or the reduced wavelength as a function of distance) and its derivatives, which provide a criterion for the validity of the method by evaluating the first of the 
discarded terms of the WKB asymptotic series \cite{Mes,LL}.

However, instead of equating the terms of the asymptotic series with the same powers of $\hbar$ to each other, it is of interest to study the case,
when the contributions from all terms of this infinite series are mutually compensated, so that the nonlinear part of the generalized Hamilton-Jacobi 
equation proportional to $\hbar^{2}$ strictly vanishes without the assumption of smallness of $\hbar^{2}$. Then the phase of the wave function is the 
classical action, whose gradient is the classical momentum of a particle. The wave properties of this particle in such a compensating (in above mentioned 
sense) medium are described by the second order partial differential equation. Introducing the index of refraction of this medium, created by the specific 
potential field, one can write the Sturm-Liouville equation equivalent to the Schr\"{o}dinger equation which contains the 
de Broglie wavelength $\lambdabar = (\hbar^{2} / 2 m E)^{1/2}$ of a particle with the mass $m$ and the energy $E$.  The Sturm-Liouville equation
has an analytical solution for both one-dimensional (1D) and three-dimensional (3D) problems. Comparison of these solutions with each other 
allows one to see the difference in wave properties of a particle in 1D and 3D spaces.

In Sect.~2, the conditions for the classicality of a quantum particle in 1D space are studied. The one-dimensional Schr\"{o}dinger equation is reduced to
the one-dimensional nonlinear Hamilton-Jacobi equation for the phase of the wave function. An explicit form of the index of refraction of the medium
created by the potential field, in which all quantum effects compensate each other, so that a quantum particle has a classical momentum, is obtained.
The field creating such a compensating medium depends not only on the coordinate, but also on the energy of the particle and its mass.
The wave function of a particle in this field is found.
In Sect.~3, the conditions for the classicality of a quantum particle in 3D space are investigated. The three-dimensional Schr\"{o}dinger equation is 
reduced to the three-dimensional nonlinear Hamilton-Jacobi equation for the phase of the wave function.  For the three-dimensional case, an explicit 
form of the index of refraction of the medium, in which quantum effects are compensated, and the form of the potential field creating such a compensating 
medium in the case of central symmetry are found. The solutions of the corresponding Sturm-Liouville equation for the wave describing a particle in 3D 
space are obtained.  Section~4 discusses the differences in behavior of particles in compensating media in 1D and 3D spaces.

\section{1D problem}\label{sec:2}

Let us consider one-dimensional motion (along the $x$-axis on the interval $x \in (-\infty, \infty)$) of a quantum particle with the mass $m$ 
and the energy $E$ in the field of the potential $V(x)$. In the case of stationary states, as is well known, 
the problem reduces to solving the Schr\"{o}dinger equation, which can be conveniently rewritten in the form
\begin{equation}\label{1} 
\psi ''(x) + \frac{1}{\lambdabar^{2}} n^{2}(x)\, \psi(x)= 0,
\end{equation}
where $\lambdabar = \hbar / p_{0} = (\hbar^{2} / 2 m E)^{1/2}$ is the de Broglie wavelength for a freely moving particle
with the  momentum $p_{0}$ and the energy $E = p_{0}^{2} / 2m$, the primes denote differentiation with respect to the variable $x$, and the function
\begin{equation}\label{2} 
n(x) = \sqrt{\frac{E - V(x)}{E}} = \frac{\lambdabar}{\lambdabar_{V}(x)},
\end{equation}
has a meaning of the index of refraction of the medium, created by the potential $V(x)$, while $\lambdabar_{V}(x) = \hbar / \sqrt{2m (E - V(x))}$ is the
wavelength for a particle moving in this potential. The solutions of Eq.~(\ref{1}) can be interpreted based on optical analogy, in which the trajectories of 
the particles are the rays.

We will look for a solution to the equation (\ref{1}) in the form 
\begin{equation}\label{3} 
\psi(x) = A(x) \exp \left(\frac{i}{\hbar} S(x) \right),
\end{equation}
where the amplitude $A$ and the phase $S$ are real functions of $x$. Substituting Eq.~(\ref{3}) into Eq.~(\ref{1})
leads to the set of two differential equations \cite{Mes,Mad,Bohm},
\begin{equation}\label{4} 
(S')^{2} - \left(\frac{\hbar}{\lambdabar}\right)^{2} n^{2} = \hbar^{2} \frac{A''}{A}, \quad
2 A' S' + A S'' = 0.
\end{equation}
The solution to the second equation is trivial: $A = C / \sqrt{S'}$, where $C = const$. As a result, we obtain the equation for a momentum 
$p \equiv S'$,
\begin{equation}\label{5} 
p^{2} - \left(\frac{\hbar}{\lambdabar}\right)^{2} n^{2} = \frac{\hbar^{2}}{2} \left[\frac{3}{2} 
\left(\frac{p'}{p} \right)^{2} - \frac{p''}{p}\right],
\end{equation} 
in which all quantum effects are contained in the nonlinear right-hand side. The equation (\ref{5}) is the generalized Hamilton-Jacobi equation.
It is exact and equivalent to Eq.~ (\ref{1}). Since $(\hbar / \lambdabar)^{2} = 2 m E$, then,
in a formal limit $\hbar \rightarrow 0$, Eq.~(\ref{5}) describes a motion of a classical particle with a momentum $p(x)$,
expressed as a function of the coordinate $x$. 

Expanding the right-hand side of Eq.~(\ref{5}) in a series in powers of $\hbar$, we get the asymptotic series (WKB method).
Restricting to a finite number of terms in a series requires careful analysis of the value of the first discarded term 
(for details, see, for example, Refs.~\cite{Mes,LL}). 

In this regard, we consider the exact solution of Eq.~(\ref{5}), which describes a quantum particle in the medium
with the index of refraction (\ref{2}). We shall assume that, when a particle is interacting with such a medium, 
the sum of contributions from all quantum corrections (in the WKB sense) strictly vanishes. In other words, these
contributions mutually compensate when summing the whole infinite series in powers of $\hbar$. At the same time, 
the ``smallness'' of $\hbar^{2}$ is not supposed.

For further analysis, it is convenient to pass to dimensionless variables $\tilde{x} = \frac{x}{\lambdabar}$ and $\tilde{p} = \hbar \lambdabar p$.
Omitting tildes in the notation of dimensionless quantities in what follows, and introducing a new unknown function $Q(x)$, such that
\begin{equation}\label{6}
p = \exp Q,
\end{equation}
we rewrite Eq.~(\ref{5}) in the form
\begin{equation}\label{7}
\exp (2 Q) - n^{2} = \frac{1}{2} \left[\frac{1}{2} (Q')^{2} -  Q'' \right],
\end{equation}
where the primes denote differentiation with respect to the dimensionless variable $x$. From Eqs.~(\ref{6}) and (\ref{7}), it follows that
the momentum (\ref{6}) will be the classical momentum of a particle for the function $Q$ satisfying the equation
\begin{equation}\label{8}
Q'' - \frac{1}{2} (Q')^{2} = 0.
\end{equation}
For the boundary condition $Q(1) = 0$, its solution is given by
\begin{equation}\label{9}
Q = \ln \frac{1}{x^{2}}.
\end{equation}
It follows from here that the classical momentum is
\begin{equation}\label{10}
p = \frac{1}{x^{2}} \quad (\mbox{or} \quad p = \frac{\hbar \lambdabar}{x^{2}}  \enskip \mbox{for dimensional quantities}),
\end{equation}
while the medium, in which such a particle is located, is characterized by the index of refraction
\begin{equation}\label{11}
n = \frac{1}{x^{2}} \quad (\mbox{or} \quad n = \left(\frac{\lambdabar}{x}\right)^{2} \enskip \mbox{for dimensional quantities}).
\end{equation}
From Eq.~(\ref{2}), it follows that the potential $V(x)$ generating such a medium,
\begin{equation}\label{12}
V(x) = E \left[1 - \left(\frac{\hbar^{2}}{2 m E} \right)^{2} \frac{1}{x^{4}} \right],
\end{equation}
is a function of not only the coordinate $x$, but also the energy $E$.
For the values $|\frac{x}{\lambdabar}| < 1$, we have $V(x) < 0$, while for $|\frac{x}{\lambdabar}| > 1$, the potential is $V(x) > 0$, and  $V (x = \lambdabar) = 0$.
Near $x \sim 0$, the potential $V(x) \sim - E \left(\frac{\lambdabar}{x} \right)^{4}$. It tends to $E$ at $x \rightarrow \infty$.

The action of a particle with a classical momentum (\ref{10}) is
\begin{equation}\label{13}
S = - \frac{1}{x} = - px \quad (\mbox{or} \quad S = - \frac{\hbar \lambdabar}{x} = - px \enskip \mbox{for dimensional quantities}).
\end{equation}
The equation (\ref{1}) reduces to the equation
\begin{equation}\label{13a}
\frac{d^{2} \psi (x)}{d x^{2}} + \frac{1}{x^{4}} \psi (x) = 0,
\end{equation}
where $x$ is expressed in units of $\lambdabar$. Its solution with the boundary condition $\psi (0) = 0$ has a form
\begin{equation}\label{14}
\psi (x) = C x \sin \left(\frac{1}{x} \right),
\end{equation}
where the constant $C = \psi (\pm \infty)$.
 
The function (\ref{14}) describes stationary states of a quantum particle with a given energy $E$ in the medium with the index of refraction (\ref{11}).
This function is shown in Fig.~1.

%Fig.~1
\begin{figure*}% figure* for wide figure, [h] [!] to change the placement
\centering
\includegraphics[width=10cm]{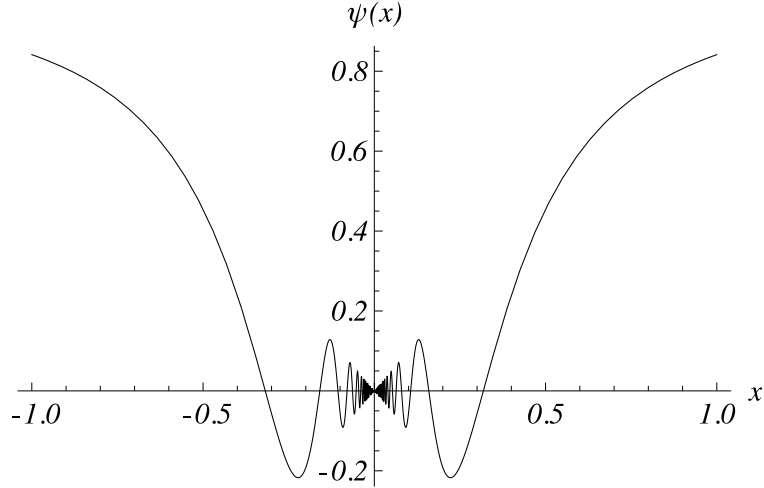}
\caption{The wave (\ref{14}) as a function of $x$ in units of the de Broglie wavelength $\lambdabar$ and for $C = 1$.}
%\label{fig:1}
\end{figure*} 

The peculiarities of the wave (\ref{14}) can be explained, if Eq.~(\ref{13a}) of the optical model is brought to the form of the standard
Schr\"{o}dinger equation. Then
\begin{equation}\label{14a}
\frac{d^{2} \psi}{d x^{2}} - U(x)\psi = -\varepsilon \psi,
\end{equation}
where
\begin{equation}\label{14b}
U(x) =  - \frac{1}{x^{4}} , \quad \varepsilon = 0,
\end{equation}
and $U$ and $\varepsilon$ are expressed in units of $E$. This equation describes the test particle with zero (binding) energy in the singular attractive 
potential $U(x)$, in which the particle can ``fall'' to the center \cite{LL}. The ``normal'' state corresponds to the particle at the origin with the infinitely large 
binding energy. The wave function  (\ref{14}) in the region $x \ll \lambdabar$ has an infinitely large number of knots and the distances between neighboring 
knots reduce as one approaches the origin. The wave (\ref{14}) occupies all the space of the variable $x$. 

Fig.~2 displays the function (\ref{14}) in the variables $\zeta = \sqrt{\frac{2 m}{\hbar^{2}}} x$ and $\sqrt{E}$ having the dimensions
[Energy]$^{-1/2}$ and [Energy]$^{1/2}$, respectively,
\begin{equation}\label{15}
\psi (x) = \psi (\zeta,\sqrt{E}) = C \sqrt{E} \zeta \sin \left(\frac{1}{\sqrt{E} \zeta} \right).
\end{equation}
%Fig.~2
\begin{figure*}% figure* for wide figure, [h] [!] to change the placement
\centering
\includegraphics[width=10cm]{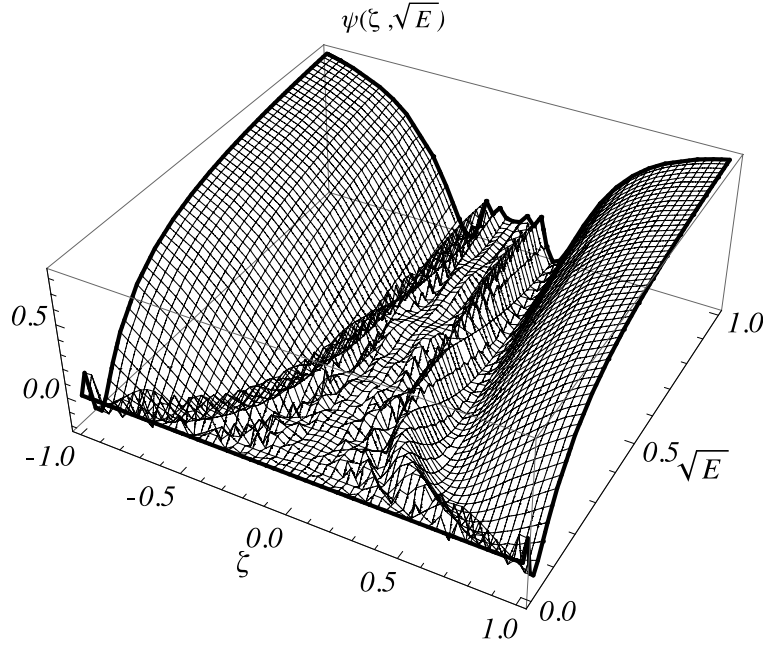}
\caption{The wave (\ref{15}) as a function of $\zeta = \sqrt{\frac{2 m}{\hbar^{2}}} x$ and $\sqrt{E}$ for $C = 1$.}
%\label{fig:2}
\end{figure*}
With increasing energy $E$, the localization region decreases for a given mass of a particle $m$. The wave shown in Fig.~1
gives a cross section of the function  $\psi (\zeta,\sqrt{E})$  corresponding to a single value of $E$.

\section{3D problem}\label{sec:3}

Let's now turn to a quantum particle with the mass $m$ and the energy $E$ being under the action of the potential $V(\bm{r})$, where $\bm{r}$ is
the radius vector drawn from the origin to a particle, which is supposed to be point-like. The stationary  Schr\"{o}dinger equation has a form
\begin{equation}\label{16}
\boldsymbol{\nabla}^2 \psi (\bm{r}) + \frac{1}{\lambdabar^{2}} n^{2}(\bm{r})  \psi (\bm{r}) = 0,
\end{equation}
where
\begin{equation}\label{17}
n (\bm{r}) = \sqrt{\frac{E - V(\bm{r})}{E}}
\end{equation}
is the index of refraction of the medium created by the potential $V(\bm{r})$. We will seek the solution of Eq.~(\ref{16}) in the form similar to (\ref{3}),
 \begin{equation}\label{18} 
\psi(\bm{r}) = A(\bm{r}) \exp \left(\frac{i}{\hbar} S(\bm{r}) \right),
\end{equation}
where the amplitude $A$ and the phase $S$ are real functions of $\bm{r}$. Substituting Eq.~(\ref{18}) into Eq.~(\ref{16}) and separating 
the real and imaginary parts, we obtain the set of two equations for $A$ and $S$ (see Ref.~\cite{Mes}),
\begin{equation}\label{19}
(\boldsymbol{\nabla} S)^{2} - \left(\frac{\hbar}{\lambdabar}\right)^{2} n^{2} = \hbar^{2} \frac{\boldsymbol{\nabla}^2 A}{A}, \quad
\boldsymbol{\nabla} \left(A^{2} \boldsymbol{\nabla} S \right) = 0.
\end{equation}
From the second equation, it follows that there exists such a vector function
\begin{equation}\label{20}
\bm{F} = A^{2} \boldsymbol{\nabla} S,
\end{equation}
whose divergence vanishes,
\begin{equation}\label{21}
\boldsymbol{\nabla} \bm{F} = 0.
\end{equation}
Since the vectors $\bm{F}$ and $\boldsymbol{\nabla} S$ are collinear, then from Eq.~(\ref{20}) it follows the condition
\begin{equation}\label{22}
A^{2} (\boldsymbol{\nabla} S)^{2} = F |\boldsymbol{\nabla} S|,
\end{equation}
where $F = |\bm{F}|$. From here, the amplitude is obtained,
\begin{equation}\label{23}
A = \sqrt{\frac{F}{|\boldsymbol{\nabla} S|}},
\end{equation}
and Eq.~(\ref{19}) reduces to the nonlinear Hamilton-Jacobi equation for the momentum $\bm{p} = \boldsymbol{\nabla} S$,
\begin{equation}\label{24}
\left(\frac{p}{\hbar}\right)^{2} - \frac{1}{\lambdabar^{2}} n^{2} = \sqrt{\frac{p}{F}} \boldsymbol{\nabla}^2 \sqrt{\frac{F}{p}},
\end{equation}
where $p = |\bm{p} |$. The right-hand side of this equation, which describes quantum effects, is equal to
\begin{equation}\label{25}
- \frac{1}{2} \left[\frac{\boldsymbol{\nabla}^2 p}{p} - \frac{3}{2} \left(\frac{\boldsymbol{\nabla} p}{p} \right)^{2} \right] 
+ \frac{1}{\sqrt{F}} \boldsymbol{\nabla}^2 \sqrt{F} + 2 \sqrt{\frac{p}{F}} \boldsymbol{\nabla} \frac{1}{\sqrt{p}} \boldsymbol{\nabla} \sqrt{F}.
\end{equation}
The modulus of the vector $\bm{F}$ in Eq.~(\ref{24}) is a function of the coordinate $\bm{r}$. Under the assumption that $\sqrt{F}$ is 
a slowly varying function of $\mathbf{r}$, such that $\boldsymbol{\nabla} \sqrt{F} \approx 0$ and $\boldsymbol{\nabla}^2 \sqrt{F} \approx 0$, 
Eq.~(\ref{24}) turns out to be independent of $F$,
\begin{equation}\label{26}
p^{2} - \left(\frac{\hbar}{\lambdabar}\right)^{2} n^{2} = \hbar^{2} \sqrt{p}\, \boldsymbol{\nabla}^2 \sqrt{\frac{1}{p}}.
\end{equation}
Similarly to a one-dimensional problem, we define a momentum $p$ as
\begin{equation}\label{27}
p(\bm{r}) = \sqrt{2 m E} \exp Q (\bm{r}).
\end{equation}
Then Eq.~(\ref{26}) will take the form simpler for further analysis 
\begin{equation}\label{28}
\exp (2 Q) - n^{2} = \frac{1}{2} \left[\frac{1}{2} (\boldsymbol{\nabla} Q)^{2} -  \boldsymbol{\nabla}^2 Q \right].
\end{equation}
Here we have passed to dimensionless quantities $\tilde{\bm{r}}  = \frac{\bm{r}}{\lambdabar}$ and $\tilde{p} = \hbar \lambdabar p$
(below we omit tildes). Quantum effects do not affect a particle with a momentum $p$, if the unknown function $Q (\bm{r})$ satisfies the equation 
\begin{equation}\label{29}
\boldsymbol{\nabla}^2 Q - \frac{1}{2}  (\boldsymbol{\nabla} Q)^{2} =  0.
\end{equation}
In the case of central symmetry, $Q = Q(r)$, Eq.~(\ref{29}) takes the form\footnotemark
\begin{equation}\label{30}
\frac{\partial^{2} Q}{\partial r^{2}} + \frac{2}{r} \frac{\partial Q}{\partial r} - \frac{1}{2} \left(\frac{\partial Q}{\partial r}\right)^{2} = 0.
\end{equation}
Its partial solution with the boundary condition $Q(1) = 0$ is following:
\begin{equation}\label{31}
Q = \ln r^{2}.
\end{equation}
The classical momentum of a quantum particle is
\begin{equation}\label{32}
p =  r^{2},
\end{equation}
and the action equals to
\begin{equation}\label{33}
S = \frac{1}{3} r^{3} = \frac{1}{3} p r.
\end{equation}
They satisfy the boundary conditions: $p(0) = 0$ and $S(0) = 0$.

The index of refraction is
\begin{equation}\label{34}
n =  r^{2}
\end{equation}
(cp. with Eq.~(\ref{11}) for the 1D problem). In the case of central symmetry, Eq.~(\ref{16}) with the index of refraction (\ref{34})
simplifies\footnotemark[\value{footnote}]
\footnotetext{Here we restrict ourselves to considering only the $s$ state.} 
\begin{equation}\label{35}
\frac{\partial^{2} \psi}{\partial r^{2}} + \frac{2}{r} \frac{\partial \psi}{\partial r} + r^{4} \psi = 0.
\end{equation}
Its solution has a form
\begin{equation}\label{36}
\psi (r) = \frac{1}{\sqrt[6]{6}} \frac{1}{\sqrt{r}} \left\{C_{1} \Gamma \left(\frac{5}{6} \right)  J_{-1/6} \left(\frac{r^{3}}{3} \right)
+ C_{2} \Gamma \left(\frac{7}{6} \right)  J_{1/6} \left(\frac{r^{3}}{3} \right)\right\},
\end{equation}
where $C_{1}$ and $C_{2}$ are constants of integration, $\Gamma$ is the Gamma function, $J_{\alpha}$ is the Bessel function of fractional order
$\alpha = \pm 1/6$. 

Near the origin, we have
\begin{equation}\label{37}
\psi (r \sim 0) \sim \frac{C_{1}}{r} +  \frac{C_{2}}{\sqrt[3]{6}}.
\end{equation}
Retaining only a regular solution that corresponds to the choice $C_{1} = 0$, we find that the function $\psi$ is constant at the origin,
\begin{equation}\label{38}
\psi (r = 0) =  \frac{C_{2}}{\sqrt[3]{6}}.
\end{equation}
At $r \rightarrow \infty$, the function $\psi$ oscillates and slowly tends to zero in accordance with the exact expression
\begin{equation}\label{39}
\psi (r) = \frac{C_{2}}{\sqrt[6]{6}} \frac{1}{\sqrt{r}}\, \Gamma \left(\frac{7}{6} \right)  J_{1/6} \left(\frac{r^{3}}{3} \right).
\end{equation}
It describes the stationary state of a quantum particle with the given energy $E$ in the medium with the index of refraction (\ref{34}) created
by the potential
\begin{equation}\label{40}
V (r) = E \left[1 - \left(\frac{2 m E}{\hbar^{2}} \right)^{2} r^{4} \right]
\end{equation}
(in physical units). It depends on the energy $E$ and is an alternating function of $r$: $V (r) < 0$ for $\left(\frac{r}{\lambdabar} \right)^{4} > 1$,
$V (r) > 0$ for $\left(\frac{r}{\lambdabar} \right)^{4} < 1$, and vanishes at $r = \lambdabar$.
%Fig.~3
\begin{figure*}[h!]% figure* for wide figure, [h] [!] to change the placement
\centering
\includegraphics[width=10cm]{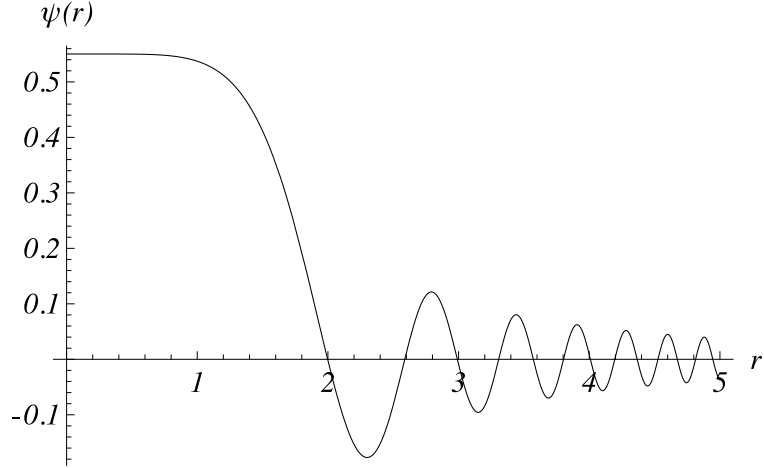}
\caption{The wave (\ref{39}) as a function of $r$ expressed in units of $\lambdabar$ and for $C_{2} = 1$.}
%\label{fig:3}
\end{figure*} 

Fig.~3 shows the function (\ref{39}), where $r$ is expressed in units of $\lambdabar$. This function is almost constant in the region $r < 1$,
where the index of refraction (\ref{34}) is small. In the region $r > 1$, the medium damps the oscillations of the wave, whose amplitude decreases as
$r^{-2}$ for $r \rightarrow \infty$. The amplitude of the corresponding wave is largest near $r = 0$, where the resonance of width $\sim 2 \lambdabar$
takes place. 

This resonance is also visible in Fig.~4, where the function (\ref{39}) is represented in Cartesian coordinates $x$ and $y$, such that 
$r = \frac{1}{\sin \theta} \sqrt{x^{2} + y^{2}}$, where $\theta$ is a polar angle. Here we put  $\theta = \frac{\pi}{2}$, when the amplitude of the wave
(\ref{39}) has a maximum value at the origin.
 %Fig.~4
\begin{figure*}[h!]% figure* for wide figure, [h] [!] to change the placement
\centering
\includegraphics[width=10cm]{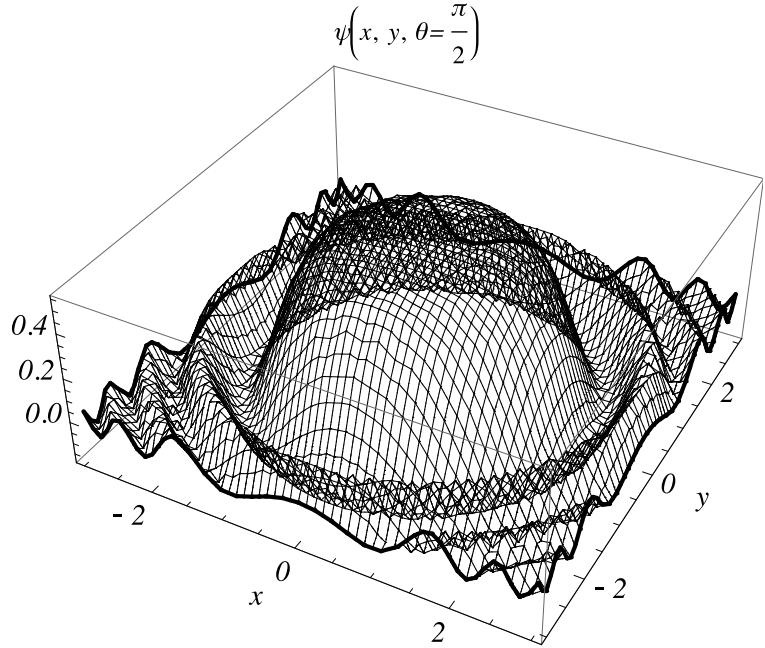}
\caption{The wave (\ref{39}) as a function of $x$ and $y$ for $C_{2} = 1$.}
%\label{fig:4}
\end{figure*} 

The peculiarities of the wave (\ref{39}) shown in Figs.~3 and 4 can be explained, if (as in the 1D case) Eq.~(\ref{35}) is rewritten in the form of the
Schr\"{o}dinger equation. As a result, we obtain the equation for the $s$ wave,
\begin{equation}\label{40a}
\frac{\partial^{2} \psi}{\partial r^{2}} + \frac{2}{r} \frac{\partial \psi}{\partial r} -U(r) \psi = - \varepsilon \psi,
\end{equation}
where
\begin{equation}\label{40b}
U(r) = - r^{4}, \quad \varepsilon = 0,
\end{equation}
and $U$ and $\varepsilon$ are expressed in units of $E$. This equation describes the test particle with zero binding energy in the attractive potential
$U(r)$. At values $r < 2$, there exists the resonance. The particle with zero energy can be found at the origin, where the attraction
of the potential is absent. At distances $r \gg 2$, the attractive action of the potential $U(r)$ damps the oscillations of the wave function, which vanishes 
only at infinity, $r = \infty$.

The behavior of the wave (\ref{39}) as a function of energy and distance from the origin is given in Fig.~5. The same variables 
$\zeta = \sqrt{\frac{2 m}{\hbar^{2}}} r$ and $\sqrt{E}$ are used as for the 1D case, but with $r \geq 0$,
\begin{equation}\label{41}
\psi (r) = \psi (\zeta, \sqrt{E}) = \frac{C_{2}}{\sqrt[6]{6}} \sqrt{\frac{1}{\zeta \sqrt{E}}}\, \Gamma \left(\frac{7}{6} \right)  J_{1/6} \left(\frac{1}{3} 
\left(\zeta \sqrt{E} \right)^{3} \right).
\end{equation}
%Fig.~5
\begin{figure*}% figure* for wide figure, [h] [!] to change the placement
\centering
\includegraphics[width=10cm]{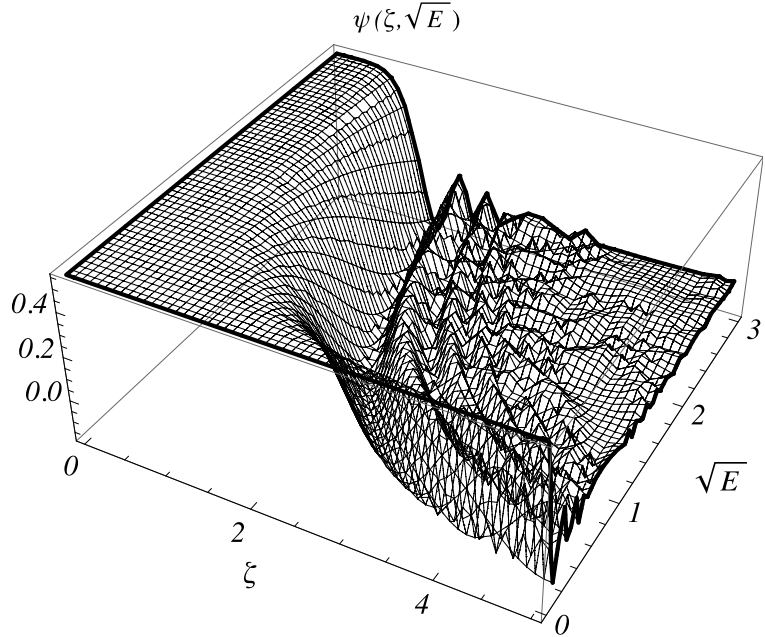}
\caption{The wave (\ref{41}) as a function of $\zeta = \sqrt{\frac{2 m}{\hbar^{2}}} r$ and $\sqrt{E}$ for $C_{2} = 1$.}
%\label{fig:5}
\end{figure*} 
With increasing $\zeta$ and/or $E$, the amplitude of this oscillating wave decreases as $\zeta^{-2}$ and/or $E^{-1}$, respectively.
Fig.~3 is a cross section of the wave $\psi (\zeta, \sqrt{E})$ corresponding to a given energy $E$.

\section{Conclusion}\label{sec:4}

In the present paper, we study the behavior of a quantum particle, when all quantum corrections (in the sense of the WKB series)
are strictly compensated. This is possible only for potential fields of a special kind (see Eqs.~(\ref{12}) and (\ref{40}) for the 1D and 3D problems, 
respectively).
The corresponding potentials depend not only on the distance from the origin, but also on the energy of the stationary state of a quantum particle.
The Schr\"{o}dinger equations (\ref{1}) and (\ref{16}) are reformulated in terms of the indices of refraction $n$ (\ref{11}) and (\ref{34}) of the media
created by the potentials $V$ (\ref{12}) and (\ref{40}), respectively. In such media, strict compensation of quantum corrections from nonlinear part of
the Hamilton-Jacobi equations  (\ref{5}) and (\ref{24}) occurs. The equations (\ref{1}) with (\ref{11}) and (\ref{16}) with (\ref{34}) 
admit exact analytical solutions in the form of Eqs.~(\ref{14}) and (\ref{39}). These solutions demonstrate differences in the wave properties of a quantum 
particle in the 1D and 3D cases. The wave (\ref{14}) is similar to that of the test particle with zero binding energy in a singular attractive potential, which
admits the ``fall'' to the center. In the 3D case, the amplitude of the wave (\ref{39}) is largest near the origin. The amplitude smoothly 
decreases in the region $r \gg \lambdabar$, where the index of refraction, given by Eq.~(\ref{34}), increases according to the quadratic law. 

The differences in the wave properties of a quantum particles in the 1D and 3D spaces can be explained by the different forms of the potentials (\ref{12}) and 
(\ref{40}) which generate the corresponding compensating media.
For the 1D problem, the potential energy (\ref{12}) takes large negative values according to $V(x) \sim - E (\lambdabar / x)^{4}$ in the region $x \ll \lambdabar$ 
ensuring the attraction of a particle to the origin. 
The potential energy (\ref{40}) of the 3D problem as a function of distance $r$ is constant, $V(r) \sim E$ for $r \ll \lambdabar$, 
and tends to negative values in accordance with $V(r) \sim - E (r / \lambdabar)^{4}$ for $r \gg \lambdabar$. As a result, the compensating medium with the 
index of refraction (\ref{34}) composes the wave structure (\ref{39}) in the form of the resonance of width $\sim 2 \lambdabar$.

Let us note that the potentials $U(x)$ and $U(r)$ of the Schr\"{o}dinger equations (\ref{14a}) and (\ref{40a}) are the asymptotics of the corresponding
potentials $V(x)$ (\ref{12}) and $V(r)$ (\ref{40}) at large distances,
\begin{equation}\label{42}
V(x) \sim - E \left(\frac{\lambdabar}{x}\right)^{4} = U(x) \quad \mbox{at} \ |x| \gg \lambdabar,
\end{equation}
\begin{equation}\label{43}
V(r) \sim - E \left(\frac{r}{\lambdabar}\right)^{4} = U(r) \quad \mbox{at} \ r \gg \lambdabar.
\end{equation}

\vskip3mm \textit{The present work was partially supported by the Program of Fundamental
Research of the Department of Physics and Astronomy of the National
Academy of Sciences of Ukraine (project No. 0117U000240).}

\end{document}